\documentclass[pra,twocolumn,showpacs]{revtex4}%
\usepackage[dvipdf]{graphicx}
\usepackage{amsmath}
\usepackage{amsfonts}
\usepackage{amssymb}%
\providecommand{\U}[1]{\protect\rule{.1in}{.1in}}
\begin{document}
\title{Optical Nanofibers for Manipulating and Probing Single-Atom Fluorescence}
\author{K. P. Nayak,$^{1}$ P. N. Melentiev,$^{1,3}$ M. Morinaga,$^{2}$ Fam Le
Kien,$^{1,\ast}$ V. I. Balykin,$^{1,3}$ and, K. Hakuta$^{1}$}
\affiliation{$^{1}$Department of Applied Physics and Chemistry, University of
Electro-Communications, Chofu, Tokyo 182-8585, Japan }
\affiliation{$^{2}$Institute for Laser Science, University of Electro-Communications,
Chofu, Tokyo 182-8585, Japan }
\affiliation{$^{3}$Institute of Spectroscopy, Troitsk, Moscow Region, 142190, Russia}
\date{\today }

\begin{abstract}
We demonstrate how optical nanofibers can be used to manipulate and probe
single-atom fluorescence. We show that fluorescence photons from a very small
number of atoms, average atom number of less than 0.1, around the nanofiber
can readily be observed through single-mode optical fiber under resonant laser
irradiation. We show also that optical nanofibers enable us to probe the van
der Waals interaction between atoms and surface with high precision by
observing the fluorescence excitation spectrum.

\end{abstract}

\maketitle

Recently, thin optical fibers with subwavelength diameters, termed as optical
nanofibers, have attracted considerable attentions. Although thin optical
fibers with diameters larger than wavelengths have been used widely for
various optical technologies, optical nanofibers are opening many new
directions in the field of optical physics. Tong \textit{et al.} developed a
method to fabricate optical nanofibers with diameters down to 50 nm and
demonstrated possible wide range of photonic applications like low loss
optical waveguiding \cite{Mazur}. Leon-Saval \textit{et al.} demonstrated
supercontinuum generation using a unique property of propagating field
confinement of nanofibers \cite{Russell}. Sumetsky \textit{et al.}
demonstrated a new type of optical resonator, micro-ring resonator, using
evanescent coupling between nanofiber coils \cite{Sumetsky}.

From a viewpoint of quantum optics, optical nanofiber may become a fascinating
work bench due to the possibility of the quantum electrodynamic effects
associated with the confinement of the fields in the guided modes. It has been
demonstrated theoretically that spontaneous emission of atoms may be strongly
enhanced around nanofibers and an appreciable amount of fluorescence photons
may be emitted into a single guided mode of the nanofibers \cite{SpCont}.
Furthermore, two distant atoms on the nanofiber surface may be entangled
through the guided mode \cite{Entangle}. Novel atom trapping schemes are also
proposed using optical nanofibers \cite{Trapping}. These possibilities may
open new approaches for manipulating single atoms and single photons, which
offer new tools for quantum information technology.

In this Letter, we experimentally demonstrate how optical nanofibers can work
for manipulating and probing single-atom fluorescence. We use laser-cooled
Cs-atoms to realize interaction time longer than the atomic spontaneous
emission lifetime. We show that fluorescence photons from atoms around the
nanofiber are measured efficiently by observing photons through the fiber
guided mode. We show also that due to the inherent nature of the nanofiber
method the fluorescence excitation spectrum strongly reflects the effect of
van der Waals interaction between atoms and nanofiber surface.

\begin{figure}[ptb]
\begin{center}
\includegraphics[width=6cm]{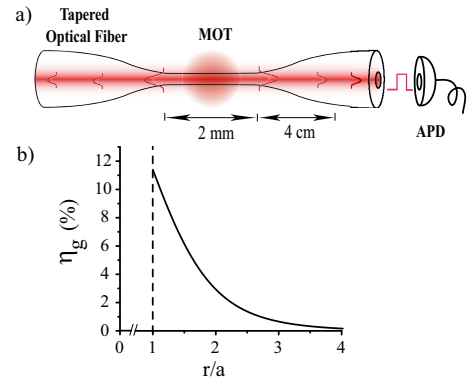}
\end{center}
\caption{(a) Conceptual diagram of the experiment. Optical nanofiber locates
at the mid of tapered optical fiber. APD denotes avalanche photodiode. (b)
Coupling efficiency of spontaneous emission into each direction of nanofiber
propagation mode, $\eta_{g}$, versus atom position $r/a$, where $r$ and $a$
are distance from nanofiber axis and radius of nanofiber, respectively.}%
\label{fig1}%
\end{figure}

Figure 1(a) shows the diagram of the present experiment. The nanofiber locates
at the mid of tapered optical fiber. A key point is the adiabatic tapering of
a single mode optical fiber so that the single-mode propagation condition can
be maintained for whole fiber length \cite{Sumetsky2}. Fluorescence photons
emitted into the guided mode are measured through the optical fiber. In Fig.
1(b) is shown the coupling efficiency of spontaneous emission into each
direction of the nanofiber propagation-mode in the evanescent region, that is
theoretically calculated based on Ref. \cite{SpCont} for a nanofiber with
$k_{0}a=1.45$, where $k_{0}$ is the free-space propagation constant and $a$ is
the radius of the nanofiber. Condition of $k_{0}a=1.45$ is the optimum
condition for fluorescence coupling into the guided mode. One can see that the
nanofiber can collect fluorescence photons very efficiently; for atoms on the
nanofiber surface, 11 \% of fluorescence photons are emitted into each side of
the guided mode, and for atoms at one radius away from the surface, still 3 \%
of photons are emitted into the guided mode.

We produce tapered fibers by heating and pulling commercial single-mode
optical fibers with cut-off wavelength of 780 nm using a fiber-coupler
production system (NTT-AT, FCI-7011). The length of the tapered region is 4 cm
on either side which ensures adiabatic tapering condition. Using this system
we make optical nanofibers with diameters from 100 to 1,000 nm. Diameters are
measured by a scanning electron microscope. The diameters are kept uniform for
2 mm along the length. Regarding the surface roughness, we have not seen any
irregularities within the resolution of 30 nm. For the present experiments, we
use the nanofibers with 400 nm diameter which satisfy the condition of
$k_{0}a=1.45$ for the D2 transition of Cs-atom.

One of the obstructions for maintaining high-transmission properties of the
optical nanofiber is dust in the air. The problem of dust is minimized by
placing the whole manufacturing unit inside a clean box and maintaining flux
of filtered air inside the box. By carrying out preparation of nanofiber under
such clean conditions high transmission of 87 \% is realized. After the
preparation, we install the nanofiber into the vacuum chamber. The vacuum
chamber is filled with dust free Ar-gas and the nanofiber is installed
horizontally into the chamber through a flange. Although the flux of Ar-gas is
maintained during installation procedure, the loss increases during the
installation and the measured transmission after installation is typically 40 \%.

We use a conventional magneto optical trap (MOT) for cold Cs-atoms. The
cooling laser is detuned 10 MHz below the closed cycle transition
($6S_{1/2}F=4\leftrightarrow6P_{3/2}F^{\prime}=5$). Each cooling and repumping
beam has an intensity of 3.3 mW/cm$^{2}$ with a beam diameter of 2 cm. The MOT
has a number density of 2$\times$10$^{10}$ atoms/cm$^{3}$ and the temperature
of trapped atoms is 200 $\mu$K. The MOT is monitored by a CCD camera during
the experiments. The MOT shape is elliptical with horizontal length of 2 mm
and vertical length of 1 mm. The position of MOT is controlled by 2 mm for
$x,y,z$-directions using the MOT quadrupole coils and other two pairs of
Helmholtz coils. After installing the nanofiber into the chamber, the MOT is
spatially overlapped with the nanofiber part of the tapered fiber using the 3
pairs of coils. By overlapping the MOT with the nanofiber, the MOT density is
reduced to 0.7$\times$10$^{10}$ atoms/cm$^{3}$, one third of the off-fiber
value, and the temperature of atoms is raised to 400 $\mu$K.

\begin{figure}[ptb]
\begin{center}
\includegraphics[width=6cm]{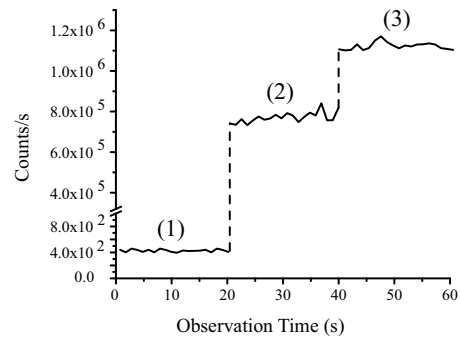}
\end{center}
\caption{Photon count through the optical fiber under three conditions; (1)
both MOT laser beams and B-fields are switched off, (2) MOT laser beams are
switched on, and (3) both MOT laser beams and B-fields are switched on.}%
\label{fig2}%
\end{figure}

First, we observe the fluorescence of MOT atoms around the nanofiber.
Fluorescence photons are measured by using an avalanche photodiode (Perkin
Elmer, SPCM-AQR/FC) connected to one end of the fiber. Signals are accumulated
and recorded on PC using a photon-counting PC-board (Hamamatsu, M8784).
Observed photon counts are plotted in Fig. 2 for three conditions. During the
first 20 seconds of observation both MOT beams and MOT magnetic-fields are
switched off and the observed 400 counts/s is due to room-light scattering and
dark counts of the detector. During the next 20 seconds MOT beams are switched
on and the observed 8$\times$10$^{5}$ counts/s corresponds to scattered light
from MOT laser beams coupled to the guided mode of nanofiber. During the final
20 seconds the MOT magnetic-fields are switched on and MOT is overlapped to
the nanofiber. An increase of 3$\times$10$^{5}$ counts/s above the scattering
background is clearly observed.

Fluorescence photon count $n_{p}$ is estimated as $n_{p}=NR\eta_{fiber}%
T\eta_{D}$, where $N$\ is effective number of atoms, $R$\ atomic scattering
rate, $\eta_{fiber}$\ averaged coupling efficiency of spontaneous emission to
the guided mode, $T$\ fiber transmission from the mid to the one end, and
$\eta_{D}$ detector quantum efficiency. $T$\ and $\eta_{D}$ are 65 \% and 45
\%, respectively. For the total laser intensity of 6$\times$3.3 mW/cm$^{2}$
and the detuning of 10 MHz, the atomic scattering rate is calculated to be
5.2$\times$10$^{6}$ s$^{-1}$ with a saturation intensity of 2.5 mW/cm$^{2}$
and spontaneous emission lifetime of 30 ns. Assuming an effective observation
volume around the nanofiber of 200 nm thickness hollow cylinder with 2 mm
length, $N$ and $\eta_{fiber}$ are estimated as $N\approx5$ and $\eta
_{fiber}\approx6$ \%, respectively. Thus, we obtain the fluorescence photon
count to be $n_{p}\approx4.6\times10^{5}$ counts/s, which is in good agreement
with the observed value.

\begin{figure}[ptb]
\begin{center}
\includegraphics[width=6cm]{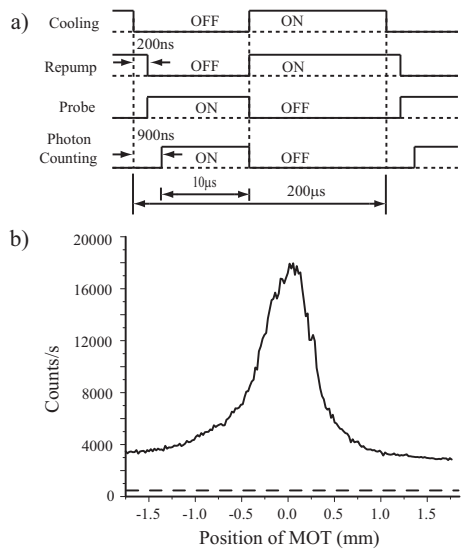}
\end{center}
\caption{(a) Time sequence for the measurements using probe laser. (b)
Observed photon count versus MOT-position. Dashed line denotes the background
count when the probe laser is off.}%
\label{fig3}%
\end{figure}

Next, we discuss the cold atom fluorescence measurements with a probe laser
beam. The observations are performed following\ the time sequence shown in
Fig. 3(a). When MOT has become ready, the cooling beam is switched off first,
and the repumping beam is switched off 200 ns later so that all the atoms can
be optically pumped into the $F=4$ hyperfine ground state. After switching off
the repumping beam, the probe beam is switched on and the photon counting is
started. The time interval between cooling-beam-off and photon-counting-on is
900 ns. The probe beam is irradiated perpendicularly to the nanofiber in a
standing wave configuration to minimize resonant photon kick on the atoms, and
the beam diameter is chosen to be 2 mm, just covering all the MOT atoms and
the uniform nanofiber region. The polarization of the probe is linear and is
perpendicular to the nanofiber axis. Photons are counted for 10 $\mu$s. Then
again the probe beam is switched off and the MOT beams are switched on for 190
$\mu$s to recollect atoms into MOT. In this way the observation is repeated at
a rate of 5 kHz and photon counts are accumulated for many cycles.

Figure 3(b) shows the observed photon count as a function of MOT position with
respect to nanofiber. MOT position is scanned across the nanofiber in vertical
direction. The probe laser frequency is tuned to the resonance of
$F=4\leftrightarrow5$ transition and the laser intensity is 3.3 mW/cm$^{2}$.
It is readily seen that scattering background has been greatly reduced
compared to the value when the MOT beams are on; the value is 2.5$\times
$10$^{3}$ counts/s. The spatial profile of MOT is clearly observed with a good
S/N ratio. The observed profile is slightly asymmetric. We have fitted the
profile with a Gaussian profile and obtained $1/e^{2}$-diameter of 1.1 mm
which well corresponds to the value of CCD measurement.

Regarding the observed fluorescence-photon-count, the peak value is 1.5
$\times$10$^{4}$ counts/s, which is about 1/20-times of that for the MOT
observation in Fig. 2. We estimate the average number of atoms in the
observation volume from the peak value using the same relation $n_{p}%
=NR\eta_{fiber}T\eta_{D}$ as for the MOT case. The number of atoms turns out
to be a very small value of $N\approx0.07$, which is 70 times smaller than the
number $N\approx5$ for the MOT case. We have measured the temporal behavior of
the peak value by setting the dark period much longer. Decay time is about 4
ms, which is well explained by the free expansion of the MOT cloud. These
observations mean that the number of atoms in the observation region has
quickly dropped to such small number during the time interval of 900 ns
between switching off the cooling beam and starting the photon-counting
\cite{outline}.

Present observations have clearly demonstrated the peculiar feature of the
nanofiber method to detect single atoms quite sensitively. The sensitivity is
due to two factors. One is the high coupling efficiency of fluorescence
photons to the guided mode. The other is that the nanofiber is immune to the
scattering from irradiating light. When we irradiate the probe beam on the
nanofiber, we readily observe bright light scattering from the nanofiber
through a CCD camera. Irradiating photon flux on the nanofiber is estimated to
be 2.3$\times$10$^{11}$ photons/s. Ideally speaking, however, such scattered
radiations cannot be coupled into the guided mode of the nanofiber, since they
are all in radiation modes and are orthogonal to the guided mode. As seen in
Fig. 3(b), we do not observe any strong scattering radiations through the
fiber; the observed scattered photon count through the fiber guided mode is
very small, 2500 counts/s, which corresponds to a scattering probability of
3.7$\times$10$^{-8}$. In this meaning, the present nanofiber works almost
ideally. Observed scattered photons might be due to some surface
irregularities of the nanofiber induced during the nanofiber production and
also by dusts on the surface. Such scattering may be reduced much more with
further technical advancement.

We measure the fluorescence excitation spectrum of atoms around the nanofiber
by scanning the probe laser frequency around the resonance ($6S_{1/2}%
F=4\leftrightarrow6P_{3/2}F^{\prime}=5$). Observed spectrum is displayed in
Fig. 4(a) for three intensities of probe laser. The observed line shape is
quite different from usual atomic line shape. Apparent power dependences
(broadenings) are not observed, implying that the spectrum is not due to an
isolated atomic spectrum, but consists of many spectral lines overlapped each
other within spontaneous linewidth. The spectrum reveals a long tail in the
red detuned side and consists of two peaks. One peak is almost on the atomic
resonance and shows a small red tail. The other peak locates around the
detuning $\Delta$ = -30 $\sim$ -50 MHz and shows a long red tail up to
$\Delta$ = -140 MHz. Obviously, these observations are attributed to van der
Waals (vdW) interaction between Cs-atom and nanofiber surface which would be
dominant for distances closer than $\lambda/2\pi$ from the surface. Since we
are observing atoms close to the nanofiber surface, appearance of the vdW
interaction should be one of the natural consequences.

\begin{figure}[ptb]
\begin{center}
\includegraphics[width=7cm]{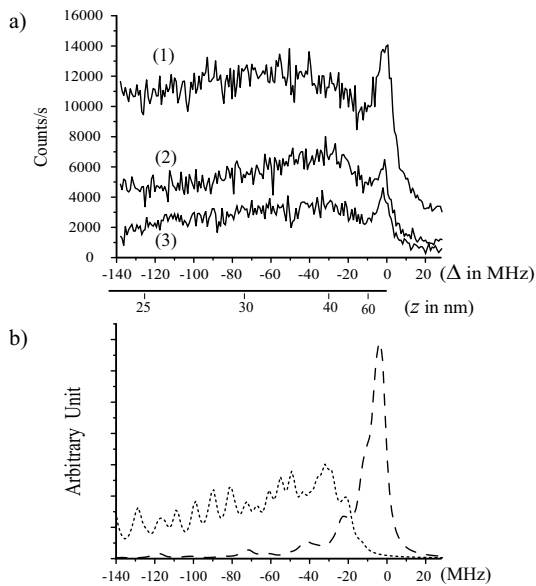}
\end{center}
\caption{Excitation spectrum versus probe laser detuning. Detuning $\Delta$ is
measured relative to atomic resonance. \ (a) Observed spectrum for three probe
laser intensities; (1) 3.2 mW/cm$^{2}$, (2) 0.64 mW/cm$^{2}$, and (3) 0.32
mW/cm$^{2}$. Detuning is calibrated to the atom distance $z$ from the surface.
(b) Calculated spectrum for the photoassociative transitions (dashed curve)
and the bound to bound transitions (dotted curve).}%
\label{fig4}%
\end{figure}

The surface induced red shift of atomic resonance is formulated as $\Delta
\nu_{\text{vdW}}\approx\nu/(k_{0}z)^{3}$, where $z$ is distance of atom from
the surface \cite{Ducloy}. Using a parameter $\nu\approx0.8$ MHz for Cs-atom
and glass surface \cite{Ducloy}, we calibrate the detuning as an atom position
from the surface as shown in Fig. 4(a). It implies that one can observe atoms
which pass through some specific distance from the nanofiber by precisely
tuning the probe laser frequency.

In order to clarify the observed spectral shape deeply, we calculate the
eigenstates of the center-of-mass motion of the atom in a close vicinity of
the nanofiber using known parameters for vdW potential for Cs-atom and silica
surface \cite{vW-potential}. Calculations are done for ground and first
excited electronic states \cite{FC-factors}. As discussed in Ref. \cite{Oria},
the eigenstates are essentially analogous to molecular vibrational states. We
calculate the Franck-Condon factors between ground and excited states and
simulate the observable spectral profiles. Regarding the highest bound states,
we included up to vibrational states with wavefunction turnig-point of
$z\approx700$ nm. Simulations are carried out for two types of transitions.
One is for the transitions from ground free-atomic states to the bound
vibrational states for the excited vdW potential. This transition corresponds
to photoassociation process; free atoms coming close to the nanofiber transit
to the upper bound molecular states, where molecule consists of nanofiber and
Cs-atom. The other is for transitions from bound ground vibrational states to
bound excited vibrational states. The simulated photoassociation spectral
profile is shown in Fig. 4(b) by a dashed plot. The profile is calculated for
atoms with velocity distribution for 400 $\mu$K. Upper vibrational states are
taken for all possible transitions. The profile is almost on the atomic line
with small red tail. Spectral profile for the ground vibrational to excited
vibrational states is shown in Fig. 4(b) by a dotted plot. We calculate the
profile assuming equal population distribution for ground vibrational states
down to a binding energy of 200 MHz. The profile reveals a long red tail with
a small offset peak position in the red side. The profile includes more than
100 spectral lines, which are overlapped each other within spontaneous
spectral width \cite{FC-factors}. We reproduce the observed spectrum with good
correspondence by adding the photoassociation and the bound to bound profiles
with adjusting the relative ratio. Thus we assign the observed red shaded peak
on the atomic line as the photoassociation process and the other broad
spectrum as due to the bound to bound vibrational transitions for atoms in the
vdW potential.

In summary, we have demonstrated how optical nanofibers can manipulate and
probe single-atom fluorescence. We have shown that very small number of atoms,
average atom number of 0.07, around the nanofiber are detected with a good S/N
ratio through single mode optical fiber under strong resonant laser
irradiation. This is essentially due to the particular feature of the optical
nanofiber around which an appreciable amount of atomic fluorescence is emitted
into the fiber guided mode. The present results imply various possible
applications, such as single-photon generation in optical fiber or EIT-based
parametric four-wave mixing \cite{EIT-NLO} using a few atoms around optical
nanofibers, which may be of importance in the context of implementing quantum
information technologies. Other than the above, the optical nanofiber may have
further potential. Regarding the spatial-resolution of atom detection, we
should mention that it may be in sub-micrometer scale, since observed
fluorescence photons are only from near surface vdW region as discussed for
the excitation spectrum. This high spatial resolution may induce some unique
applications in atom optics. The inherent nature of the nanofiber method to
detect vdW interaction between atoms and surface may open new possible
directions to investigate atom-surface interaction with high precision. The
optical nanofiber method can naturally be extended to other systems than
atoms, like molecules or quantum-dots.

This work was carried out under the 21st Century COE program on Coherent
Optical Science.


\begin{thebibliography}{999}
\bibitem {outline}Detailed dynamics is discussed elsewhere. The outline is as
follows. Stationary atom number in the observation region is determined by a
balance between atom loading to the region and loss in the region. Atoms in
the region may finally be adsorbed to the nanofiber surface or reflected back
from the surface. Both processes act as the loss processes in the observation
region. Regarding the atom loading, the rate may be much higher with MOT than
without MOT beams, because atoms in MOT are forced to direct toward the trap
center due to friction force through cooling beam, whereas atoms without MOT
beams are free and simply enter into the observation region just through
diffusion process.
\end{thebibliography}
\end{document}